\documentclass[]{elsart}

\usepackage{epsfig}

\newcommand{\megp}{$\mu^+\rightarrow\mathrm{e}^+\gamma$\ }%

\newcommand{\npe}{$N_\mathrm{pe}$}
\newcommand{\attlen}{$\lambda_\mathrm{att}$}
\newcommand{\abslen}{$\lambda_\mathrm{abs}$}
\newcommand{\scalen}{$\lambda_\mathrm{sca}$}
\newcommand{\raylen}{$\lambda_\mathrm{R}$}

\begin{document}

\begin{flushright}
{\large UT-ICEPP 04-02}
\end{flushright}

\begin{frontmatter}

\title{Absorption of Scintillation Light in a 100~$\ell$ Liquid Xenon
$\gamma$ Ray Detector and Expected Detector Performance}

\author[pisa]{A. Baldini},
\author[pisa]{C. Bemporad},
\author[pisa]{F. Cei},
\author[waseda]{T. Doke},
\author[pisa]{M. Grassi},
\author[binp]{A. A. Grebenuk},
\author[binp]{D. N. Grigoriev},
\author[kek]{T. Haruyama},
\author[kek]{K. Kasami},
\author[waseda]{J. Kikuchi},
\author[kek]{A. Maki},
\author[tokyo]{T. Mashimo},
\author[tokyo]{S. Mihara},
\author[tokyo]{T. Mitsuhashi},
\author[tokyo]{T. Mori},
\author[pisa]{D. Nicol\`o},
\author[tokyo]{H. Nishiguchi},
\author[tokyo]{W. Ootani},
\author[tokyo]{K. Ozone\corauthref{cor}},
\corauth[cor]{Corresponding author. Tel.: +81-3-3815-8384; fax:
+81-3-3814-8806.}
\author[pisa]{A. Papa},
\author[pisa]{R. Pazzi},
\author[psi]{S. Ritt},
\author[tokyo]{R. Sawada},
\author[pisa]{F. Sergiampietri},
\author[pisa]{G. Signorelli},
\author[waseda]{S. Suzuki},
\author[waseda]{K. Terasawa},
\author[waseda]{M. Yamashita},
\author[tokyo]{S. Yamashita},
\author[waseda]{T. Yoshimura},
\author[binp]{Yu. Yuri}
\address[waseda]{Advanced Research Institute for Science and
Engineering, Waseda University, Tokyo, Japan}
\address[binp]{Budker Institute of Nuclear Physics, Novosibirsk, Russia}
\address[kek]{High Energy Accelerator Research
Organization~(KEK), Tsukuba, Japan}
\address[pisa]{INFN Pisa, University and Scuola Normale Superiore di
Pisa, Italy}
\address[tokyo]{International Center for Elementary Particle Physics and 
Department of Physics, University of Tokyo, Tokyo, Japan}
\address[psi]{Paul Scherrer Institut, Villigen, Switzerland}
\date{\today}
\begin{abstract}

An 800~$\ell$ liquid xenon scintillation $\gamma$ ray detector is being
developed for the MEG experiment which will search for \megp decay
at the Paul Scherrer Institut.
Absorption of scintillation light of xenon
by impurities might possibly limit the performance of such a detector.
We used a 100~$\ell$ prototype with an active volume of
$372\times 372\times 496$~mm$^3$ to study the scintillation
light absorption.
We have developed a method to evaluate the light absorption, separately
from elastic scattering of light, by measuring
cosmic rays and $\alpha$ sources.
By using a suitable purification technique,
an absorption length longer than 100~cm has been achieved.
The effects of the light absorption on the energy resolution
are estimated by Monte Carlo simulation.
\end{abstract}
\end{frontmatter}

\section{Introduction}
\label{sec:intro}

A liquid xenon~(LXe) scintillation $\gamma$ ray detector is being
developed for the MEG experiment which will search for \megp decay
at the Paul Scherrer Institut~\cite{proposal}.
The MEG experiment aims to improve the sensitivity by at least
two orders of magnitude over the previous experiment~\cite{mega} and
to reach a \megp branching ratio that is predicted by
theories of supersymmetric grand unification~\cite{susygut} and
supersymmetric theories with right-handed neutrinos~\cite{RHneutrino}.

The LXe detector is an indispensable component of the experiment
that enables to achieve such a superior sensitivity with its excellent
performance in $\gamma$ ray measurement.
It will have an active LXe volume of 800~$\ell$ surrounded
by 800 photomultiplier tubes (PMTs) viewing inside~\cite{proposal}.
The detector utilizes only scintillation light without any attempt to
measure ionization.
The principle of the detector has been successfully demonstrated
by a small prototype with an active LXe volume
of 2.34~$\ell$ surrounded by 32 PMTs~\cite{exp32}.

However, since such a large LXe detector has never been built,
nobody has verified
long distance transparency of LXe for the vacuum ultra violet (VUV)
scintillation light of xenon.
While LXe itself should never absorb the scintillation,
practically unavoidable impurities might deteriorate
the transparency and thus the performance of the detector.

We have recently built
a 100~$\ell$ prototype with an active volume of
$372\times 372\times 496$~mm$^3$ (69~$\ell$) surrounded by 228 PMTs
to gain practical experiences in operating such a device
and to prove its excellent performance.
In this paper we describe how this prototype has been used
to evaluate the absorption of the scintillation light in LXe.
Based on the obtained result,
the effects of the light absorption on the energy resolution
are estimated by Monte Carlo simulation.

\section{Optical properties of LXe}

We discuss here the optical properties of LXe that are of interest
for a scintillation detector.
The main optical properties of LXe are
listed in Table~\ref{tab:xeprop}.

\begin{table}[!htb]
\caption{Optical properties of LXe.}
\label{tab:xeprop}
\begin{center}
\begin{tabular}{llll} \hline
Material Properties & Value \& Unit & Ref.\\ \hline
Refractive index at scintillation wavelength
  & $1.6 - 1.7$ & \cite{solovov,barkov,lan}\\
Peak scintillation wavelength & $178$~nm & \cite{jort1,jort2} \\
Scintillation spectral width (FWHM) & $\sim 14$~nm & \cite{jort1,jort2} \\
Absorption length $\lambda_{\rm abs}$ & $>100$~cm & present work\\
Scattering length $\lambda_{\rm sca}$
  & 29 - 50~cm & \cite{solovov,lan,rayl1,rayl2,rayl3}\\
\hline
\end{tabular}
\end{center}
\end{table}

Light attenuation in LXe can be described by the attenuation length,
$\lambda_{\rm att}$, as defined in the equation,
  $I(x)= I_{0}~ e^{-x/{\lambda_{\rm att}}}$.
The attenuation length consists of two separate components,
the absorption length, \abslen, describing real absorption and
loss of photons, and
the scattering length, \scalen, that represents
elastic scattering of photons without any loss.
For the elementary photon scattering they are related by:
$1/\lambda_{\rm att}=1/$\abslen$+1/$\scalen.
For an actual photon beam one must take into account the fact that
photons can be forward scattered.
In our case the elastic scattering is dominated by Rayleigh scattering,
therefore \scalen\ may be regarded as the Rayleigh scattering length,
\raylen.
The knowledge of refraction index $n$ in the
region of the xenon VUV light emission is also relevant.

The most important among these parameters for the detector performance
is the absorption length \abslen,
since the energy measurement relies on the total number of scintillation
photons detected by the PMTs that surround and view the LXe volume
from all sides and is therefore
not affected by photon scattering.

The experimental knowledge of these quantities, $\lambda_{\rm att}$,
\abslen, \scalen~ and $n$ for pure LXe is poor.
Especially no measurement of \abslen~ has been made before.
There are also some discrepancies among the available experimental data,
which might partly be explained by the degree of purity of the LXe.
As we shall show later,
small amounts of VUV absorbing molecules like H$_{2}$O or O$_2$,
at the level of a few parts per million,
can dramatically change LXe optical parameters.
Thus a reliable determination of optical parameters necessarily requires
monitoring the level and the stability of the xenon purity.
In addition a size of the LXe volume comparable to
the absorption/scattering lengths is needed
to make a reasonable measurement of these lengths.

Considering the scintillation mechanism of LXe through
the excimer state Xe$^*_2$~\cite{xeprocess},
absorption in pure LXe is
improbable, i.e. \abslen$\sim\infty$; any absorption is
thus caused by VUV absorbing impurities.
In this paper we report the first measurement of \abslen\ using our prototype
and present a significant lower limit.

In contrast to the situation for LXe,
better information is available for
gaseous xenon (GXe) at various pressures both for visible and VUV
light~\cite{angela1,angela2,angela3}.
One can then examine if optical properties for LXe can be derived from
those for GXe, although this might imply a daring extrapolation
in terms of density.
For a non-polar gas like xenon, however,
this extrapolation proves to be valid~\cite{lan}.
One has to ascertain up to which value of photon energies
the extrapolation maintains its validity. This point was further
investigated by us~\cite{xeoptical};
the extrapolation is reliable up to a photon energy of 8.1~eV.
At higher energies and closer to
the first xenon absorption line at 8.3~eV
the extrapolation is no longer valid.
Our prediction for the value of the LXe refractive index
at its emission line (7~eV corresponding to the wavelength of 178~nm)
is $n = 1.65\pm 0.03$.
Most of the information obtained for GXe in various physical conditions
can be used for reliably predicting other LXe optical
properties~\cite{xeoptical}.
Here we restrict the discussion to the relation between two quantities:
$n$ and \raylen.

For a dense fluid like LXe \raylen\ depends on density and
temperature fluctuations of the medium, according to the Einstein's
expression~\cite{landau}:
\begin{equation}
{1 \over \lambda_\mathrm{R}}  =  \frac{\omega^{4}}{6 \pi c^{4}}
  \left[  K T \rho^{2} \kappa_{T}
            \left(\frac{\partial \epsilon}{\partial \rho}\right)^{2}_{T}
        + \frac{K T^{2}}{\rho c_{v}}
            \left(\frac{\partial \epsilon}{\partial T}\right)^{2}_{\rho}
  \right]
  \label{eqn:h1}
\end{equation}
where $\epsilon$ is the dielectric constant,
$\kappa_{T}$ is the isothermal compressibility, $c_{v}$ is the specific
heat at constant volume and $K$ is the Boltzmann's constant.

Since xenon is a non-polar fluid, the second part of Eq.~\ref{eqn:h1}
comes out to be negligible \cite{lan,sin69,hohm}.
The derivative appearing in the first part of Eq.~\ref{eqn:h1} can be
computed from the Clausius-Mossotti equation:
\begin{equation}
\frac{\epsilon(\omega)-1}{\epsilon(\omega)+2} =
  \frac{4 \pi}{3} \frac{N_{A} \alpha(\omega) \rho}{M},
\label{a1}
\end{equation}
where $N_{A}$ is the Avogadro's number,
$\alpha(\omega)$ is the molecular polarizability
and $M$ is the molecular weight.
The Einstein's equation reduces then to:
\begin{equation}
{1 \over \lambda_\mathrm{R}} =
  \frac{\omega^{4}}{6 \pi c^{4}}
  \left[K T  \kappa_{T}\frac{(n^2-1)^{2}(n^2+2)^{2}}{9}\right].
\label{eqn:h3}
\end{equation}
This equation establishes therefore a useful relation between the index
of refraction in pure LXe and the Rayleigh scattering length.

\section{The 100~$\ell$ prototype}

\begin{figure}[!htb]
\begin{center}
\epsfig{file=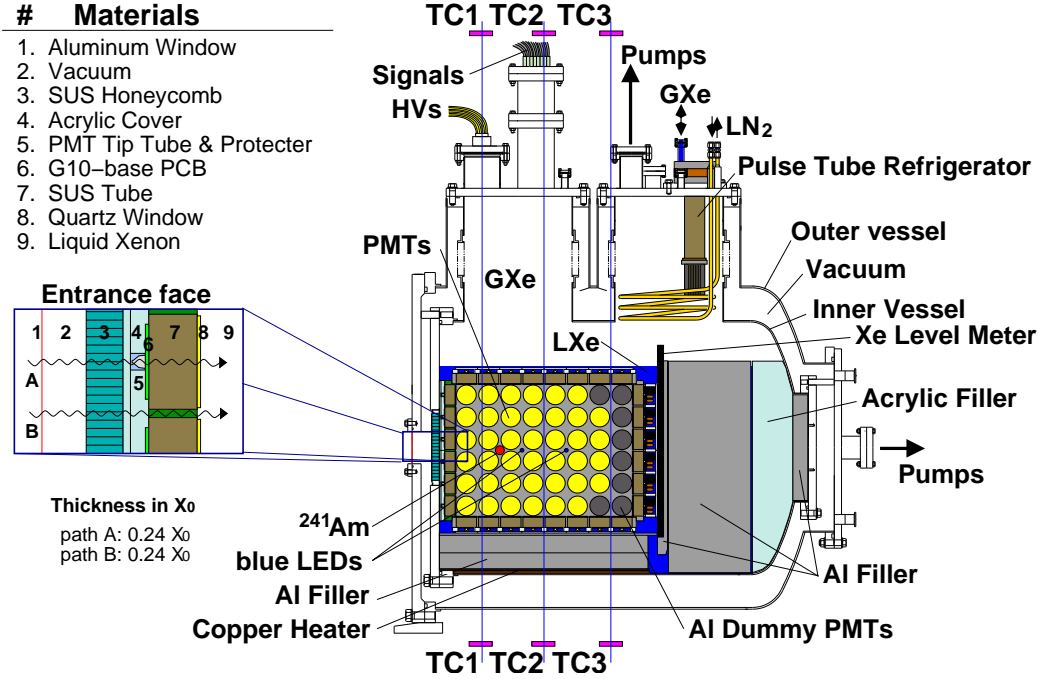,width=14cm}
\end{center}
\caption{A schematic drawing of the 100~$\ell$ prototype.
Shown on the left is a zoomed view of the front materials
the $\gamma$ rays traverse.}
\label{fig:lpsetup}
\end{figure}

A schematic view of the prototype detector is shown in
Fig.~\ref{fig:lpsetup}.
It has an active volume of
$372\times372\times496~\mathrm{mm}^3$~(69~$\ell$) viewed from all sides
by 228 PMTs assembled into a rectangular shape.
The cryostat consists of thermal insulated vessels
equipped with a pulse tube refrigerator~\cite{refrige,aisin} and
a liquid nitrogen cooling pipe.
Several sensors are used inside the vessels
for monitoring temperatures and pressures.
A signal feedthrough
that was originally developed for
the ATLAS liquid argon calorimeter~\cite{atlas}
is installed on the flange of the vessel.

To minimize $\gamma$ ray interactions before
the active region of the detector,
a thin aluminum plate and a honeycomb window made of stainless steel
are used as an entrance window.
Including the PMTs and their holders,
the front materials amount to a total thickness of 0.24~$\mathrm{X}_0$,
as summarized in Fig.~\ref{fig:lpsetup}.

All the PMTs are immersed in LXe in order to maximize the direct
light collection.
The PMTs (HAMAMATSU R6041Q~\cite{pmt}) operate at the LXe
temperature and stand up to 0.3~MPa pressure.
They have been specially developed in
cooperation with Hamamatsu Photonics K. K.
Their properties are summarized in Table~\ref{tab:pmt}.
The short axial length of 32~mm is realized by
the adoption of metal channel dynode structure.
A synthetic quartz window is used to allow for the VUV
scintillation light~\cite{jort2,wvalue2}.
The circuit elements that operate even at the liquid nitrogen temperature
were carefully selected for the voltage dividers.

The detector is equipped with 8 blue LEDs and
four $^{241}$Am $\alpha$ sources,
two LEDs and one $\alpha$ source on each lateral face,
for PMT calibration.
Three pairs of trigger counters~(TC1, TC2, and TC3) are placed
above and below the vessel
to select cosmic ray muons traversing the LXe
for various tests.

% PMT properties
\begin{table}[!htb]
\caption{Properties of the HAMAMATSU R6041Q PMT.
Note that ``quantum efficiency'' (QE) includes the collection efficiency
at the first dynode and the transmission through the quartz window.}
\label{tab:pmt}
\begin{center}
\begin{tabular}{|l|c|}
\hline
Diameter&$\phi$~57 mm\\
\hline
Photo-Cathode Material& Rb-Cs-Sb\\
\hline
Active Area& $\phi$~46~mm\\
\hline
QE at 165~K&6~\% typ.\\
\hline
Dynode Type&Metal channel\\
\hline
Number of Stages&12\\
\hline
Supplied H.V.&900~V typ.\\
\hline
Gain& $10^6$ typ.\\
\hline
Thickness~(center)& 0.20~$\mathrm{X}_0$\\
\hline
\end{tabular}
\end{center}
\end{table}

\section{Detector operations}

The detector requires a total of about 100~$\ell$ of LXe.
Some preparations are needed to bring it into operational
conditions.
Once in operation, the gains of the PMTs are measured with the blue LEDs
and adjusted by changing the high voltage.
The effective quantum efficiencies (QE) are estimated by measuring
scintillation spectra of the $\alpha$ sources in GXe at 170~K,
close to the liquid temperature.

\subsection{Liquefaction process}

Both the inner and outer vessels are initially evacuated.
While being evacuated, the inner vessel is baked at the rather low
temperature of $70~{}^\circ$C because of the PMTs and acrylics
inside the vessel.
After a continuous evacuation for about ten days, the inner pressure
reaches $10^{-3}$~Pa.

In order to cool the inner vessel down
to 165~K prior to xenon liquefaction (pre-cooling),
it is first filled with GXe at 0.2~MPa
and then cooled for a whole day by the combined action of
the pulse tube refrigerator and liquid nitrogen
flowing through a coiled stainless steel pipe in the inner vessel.
The refrigerator, operating quietly
with no moving parts near the cooling head,
does not require frequent interruptions for maintenance~\cite{aisin}.

When the inner vessel is sufficiently pre-cooled,
xenon is finally liquefied by using liquid nitrogen and the refrigerator.
GXe flows through a gas purifier (SAES Getter\cite{saes})
and molecular filters (Oxisorbs~\cite{oxisorb})
before entering the vessel and getting liquefied.
The gas purifier absorbs various contaminants such as
$\mathrm{H}_2\mathrm{O}, \mathrm{O}_2,
\mathrm{CO}_2, \mathrm{N}_2,$ and $\mathrm{CH}_4$ down to the ppb level.
The molecular filters act as an additional oxygen trap.
It usually takes two days to liquefy 100~$\ell$ of LXe.
The vapor pressure in the inner vessel is continuously monitored to
regulate the flow of liquid nitrogen.

After liquefaction,
LXe is maintained at 168~K and 0.13~MPa mainly by the refrigerator.
For the measurement described later in this paper, the detector was kept
in operation continuously for approximately 2000~hours.

After detector operation xenon is recovered to a storage tank.
The refrigerator is turned off, the outer vessel is filled with
nitrogen gas of room temperature,
and a heater under the PMT holder is switched on
to accelerate LXe evaporation.
The storage tank is cooled down by liquid nitrogen in order to re-condense
the xenon flowing from the detector vessel.
Recovery and warming up of the cryostat take two and four days,
respectively.

\subsection{PMT calibration}
\label{subsec:calib}
A precise knowledge of PMT gains and quantum efficiencies is necessary
to have an excellent energy resolution.
The blue LEDs and the $^{241}$Am $\alpha$ sources placed inside
the detector are used for the calibration of the PMTs.

\subsubsection{Gain monitoring and adjustment}
\label{subsubsec:gain}
\begin{figure}[!htb]
\begin{center}
\epsfig{file=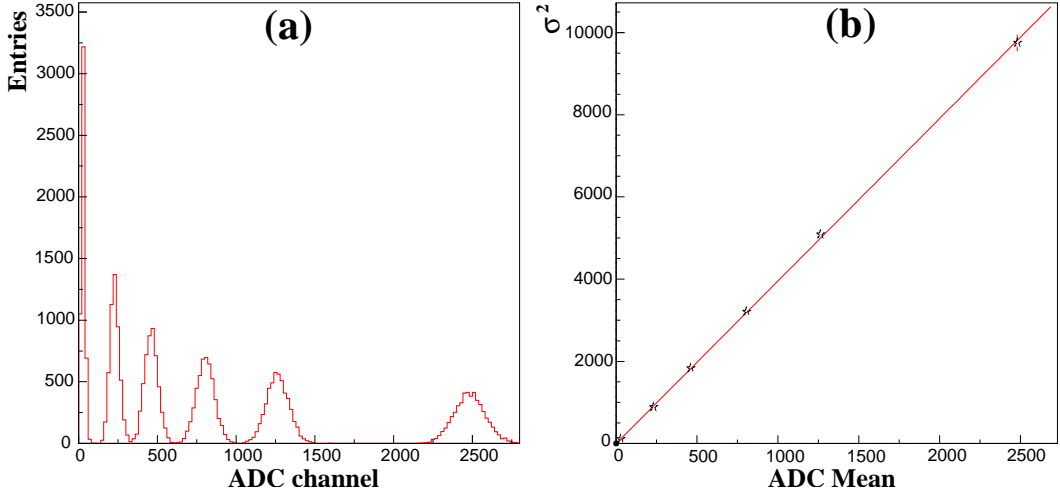,width=14cm}
\end{center}
\caption{
(a) A typical ADC spectrum of a PMT in one of the LED gain calibration runs.
(b) The relation between $\sigma^2$s and the mean channels $M$.
The pedestal mean is already subtracted and
$\sigma^2$ at $M = 0$ corresponds to $\sigma^2_0$.}
\label{fig:ledcal}
\end{figure}

During data acquisition, that lasted over 2000 hours,
the gain of each PMT was monitored twice a day by flashing, at a time,
a pair of LEDs facing each other
at 100~Hz by LED drivers~(CAEN C529~\cite{caen}) at several different intensities.
A typical ADC spectrum of the PMT outputs in one of these LED runs is
shown in Fig.~\ref{fig:ledcal}~(a).
Assuming negligible fluctuations of the LEDs intensity,
the gain $g$ can be given by the following equation:
\begin{equation}
g = \frac{c\sigma^2}{eM}
\label{eq:gain1}
\end{equation}
where $c$ is the charge per ADC channel~($200~\mathrm{fC/ch}$),
$\sigma$ and $M$ are the standard deviation and the mean of ADC
spectrum, respectively,
and $e$ is the electron charge magnitude.
Here it is assumed that the number of photoelectrons~($N_{pe}$) observed on a
PMT is reasonably large so that the spectrum can be regarded as Gaussian.
In practice we have to consider a contribution from the
pedestal as in the following equation:
\begin{equation}
\sigma^2=g\frac{e}c(M-M_0)+\sigma^2_0,
\label{eq:gain2}
\end{equation}
where $M_0$ and $\sigma_0$ are the mean and the standard deviation of
the pedestal.
Fig.~\ref{fig:ledcal}~(b) shows an example of the linear relation
between $\sigma^2$ and $M$.  The gain of the PMT is evaluated by
fitting the data with Eq.~\ref{eq:gain2}.
To evaluate reliability of the obtained results,
the procedure was repeated both by using different LED pairs and by using
the same pair in several consecutive runs.
The results were reproduced within 0.9~\% in FWHM,
which gives a negligible contribution to the energy resolution
of the detector.

\subsubsection{Determination of quantum efficiencies}
\label{subsubsec:qe}
Quantum efficiencies (QEs) of the PMTs
depend on the temperature and the light wavelength
and must be evaluated in the operational conditions.
They were evaluated by using the $\alpha$ sources
with 0.13~MPa GXe at 170~K.
The use of gaseous xenon is necessary to avoid scattering and possible
absorption of the scintillation photons,
which are more prominent in liquid phase.
The scintillation spectra in liquid and gaseous phases are not much
different~\cite{jort1,lumine}.

\begin{figure}[!htb]
\begin{center}
\epsfig{file=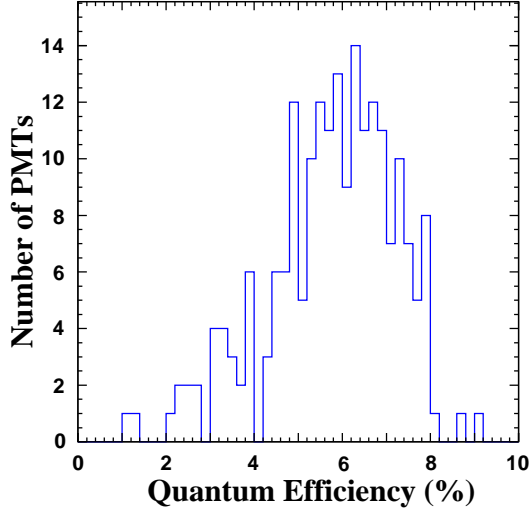,width=7cm}
\end{center}
\caption{The Distribution of QEs evaluated in GXe.}
\label{fig:qe}
\end{figure}

QEs for individual PMTs were evaluated by comparing the measured spectra
with a GEANT3 simulation~\cite{g3}.
The obtained QE distribution for all the PMTs is shown in Fig.~\ref{fig:qe}.
The low values and the broad spread of QEs are thought to be due to
the increased photo-cathode resistivity at low temperature and
the difficult evaporation procedure of the photo-cathode.\footnote{
New types of PMTs have been recently developed
by coating Al strips on the quartz window
to reduce the resistivity and by using more standard photo-cathode
material to ease the evaporation procedure.
A preliminary measurement shows their QEs are about 15--20~\%.}

The $\alpha$ sources were also used for monitoring the stability
of the PMTs during the data taking in liquid phase.
It is found that the PMT outputs stabilize in about 50 hours
after the liquefaction and are stable within 0.5~\% thereafter.

\section{Xenon purification}
\label{sec:purify}

LXe should in principle be transparent to its own scintillation light
thanks to the scintillation mechanism through the excimer state
$\mathrm{Xe}^\mathrm{*}_2$~\cite{xeprocess}.
However possible contaminants in LXe, such as water and oxygen at ppm level,
considerably absorb scintillation light~\cite{h2oabs,o2abs}.

\begin{figure}[!htb]
\begin{center}
\epsfig{file=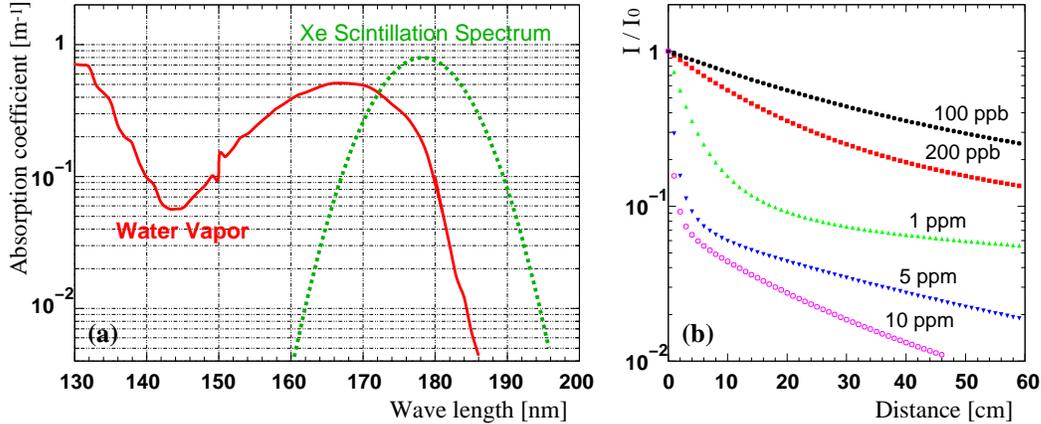,width=14cm}
\end{center}
\caption{
(a) Absorption coefficient for 1~ppm water vapor.
Superimposed is the xenon scintillation spectrum.
(b) Scintillation light intensity as a function of the distance
from the light source for various concentrations of water in LXe.}
\label{fig:contami}
\end{figure}

In Fig.~\ref{fig:contami}~(a) the absorption coefficient for VUV
light is shown for 1 ppm contamination of water vapor.
The absorption spectra of water and oxygen largely overlap with
the xenon scintillation spectrum.
Given these absorption coefficients and neglecting the scattering
(\abslen$<$\raylen), we calculated the light intensity as a function
of the distance from the light source for various concentrations of
the contaminant.  The result is shown in
Fig.~\ref{fig:contami}~(b) for water.
Since water tends to absorb light with shorter wavelengths,
only a component with longer wavelengths survives for a long distance.
This might explain the discrepancies among the measurements of the LXe
refractive index $n$, as $n$ varies rapidly as a function of the
wavelength in the vicinity of the scintillation wavelength.

As we already noted, xenon is well purified before filling the
detector vessel, but
some inner components of the prototype are made of G10 and acrylic
that are known to absorb and desorb water.
During the initial stage of the prototype operation,
a strong absorption of scintillation light was observed.
After systematic studies on the residual gas
by means of mass spectroscopy and build-up tests,
we concluded that water at the ppm level seemed to be desorbed from the
detector material into LXe.

\begin{figure}[!htb]
\begin{center}
\epsfig{file=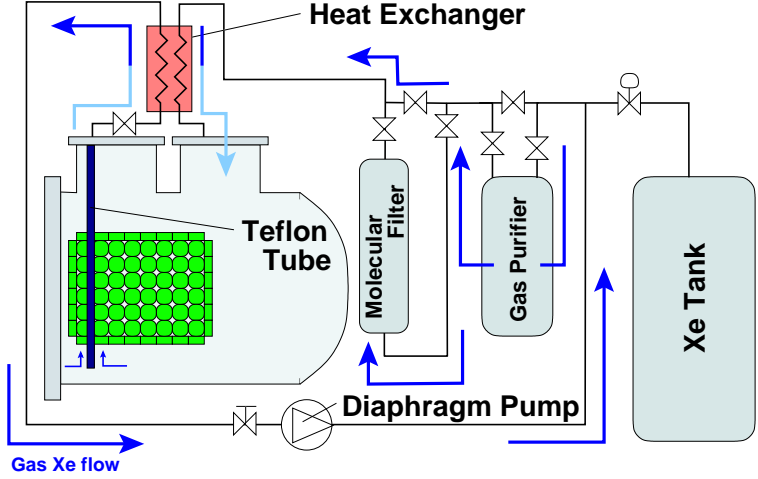,width=10cm}
\end{center}
\caption{The circulation and purification system of xenon.}
\label{fig:circle}
\end{figure}

We therefore introduced a circulation-purification system,
as schematically shown in Fig.~\ref{fig:circle}, to remove the water.
In this system, xenon,
taken from the bottom of the vessel through the Teflon tube
and evaporated by the heat exchanger,
is pumped by a diaphragm pump
and circulated through the gas purifiers and the molecular filters,
and is finally condensed back into the detector.
The flow rate of GXe is about 500~$\mathrm{cm}^3$/min, hence
the whole volume could be circulated in a few month time.
We also carried out various tests to study the purification process,
such as stopping or changing the flow rate and bypassing the purifiers
or the filters.

\section{Absorption length estimate}
\label{sec:abs}

Purification was continuously performed for over 1200 hours.
To evaluate and monitor light absorption in LXe
separately from Rayleigh scattering during purification,
we used cosmic ray muons as well as the $\alpha$ sources.

\begin{figure}[!htb]
\begin{center}
\epsfig{file=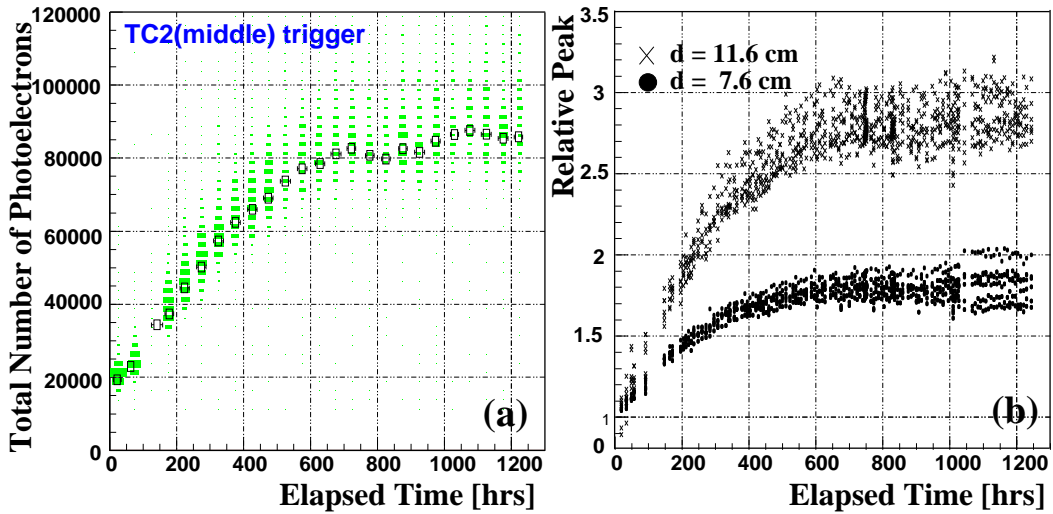,width=14cm}
\end{center}
\caption{(a) The total number of photoelectrons collected by all the PMTs
for cosmic ray events as a function of time.
(b) The relative changes in the $\alpha$ peaks of the PMTs located at
7.6~cm and 11.6~cm from the $\alpha$ sources.}
\label{fig:crnpe}
\end{figure}

The total number of photoelectrons collected by all the PMTs (\npe)
for each cosmic ray event, being sensitive only to the loss of the
scintillation photons, is a good measure of the light absorption.
In Fig.~\ref{fig:crnpe} (a), \npe\ is plotted as a function of time.
It increased by a factor four in about one month
(700~hours) and almost saturated. A comparison with a Monte Carlo simulation
indicates that \abslen\ increased from $\sim 10$~cm to above 1~m.

In Fig.~\ref{fig:crnpe} (b)
the relative changes in the $\alpha$ peaks of the PMTs located
at certain distances (7.6~cm and 11.6~cm) from the $\alpha$ sources
are plotted.
For the PMTs at a longer distance,
the PMT outputs increased much more significantly and saturated slightly
later, just as expected for the light absorption in LXe.

We first made a crude estimate of the Rayleigh scattering length
by using the PMTs located on the same face as the $\alpha$ source;
these PMTs can not directly see the scintillation light
from the $\alpha$ particles but only the scattered light.
Our data prefer values of \raylen $=$ 40 - 50~cm,
which are consistent with the numbers currently
available in the literature~\cite{solovov,lan,rayl1,rayl2,rayl3}.

\begin{figure}[!hbt]
\begin{minipage}{0.5\textwidth}
  \centering\epsfig{file=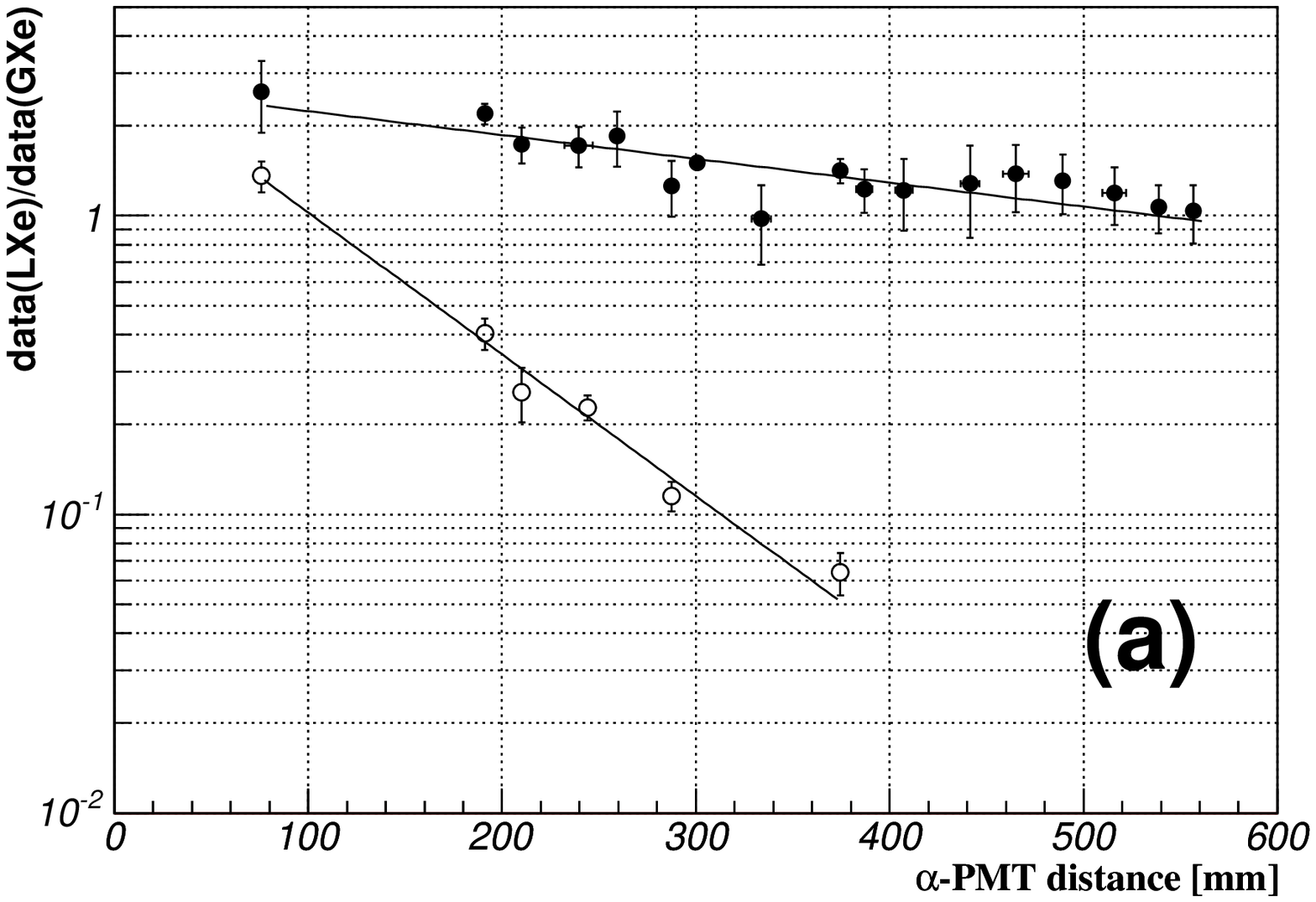,width=\textwidth}
\end{minipage}
\begin{minipage}{0.5\textwidth}
  \centering\epsfig{file=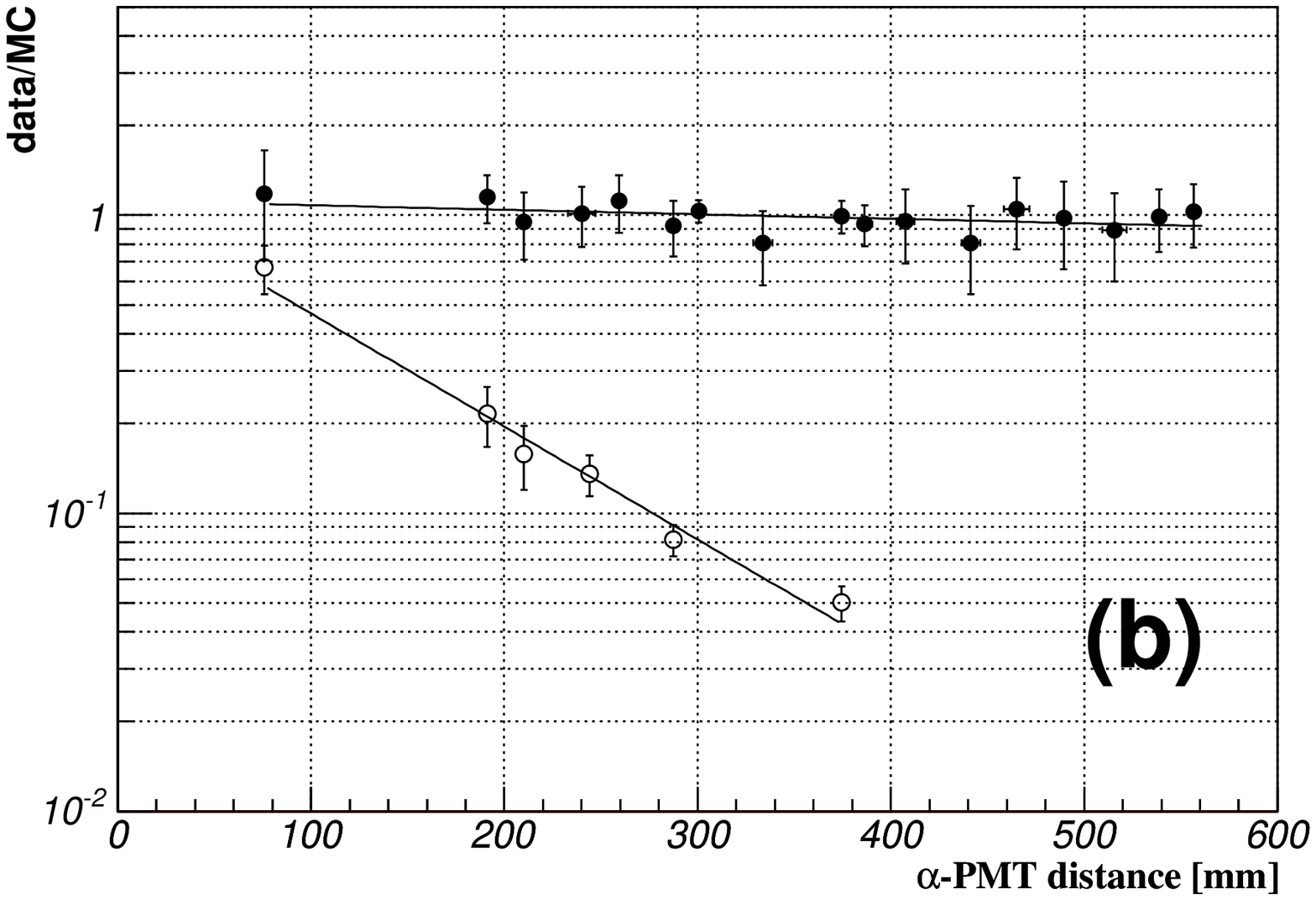,width=\textwidth}
\end{minipage}
\caption{The PMT outputs normalized either
to those in the GXe data (a) or
to the MC simulation without absorption (b)
are plotted as a function of the source-PMT distance
at the beginning (open circles) and after purification
(closed circles). The solid lines are fitted results.}
\label{fig:abs}
\end{figure}

To evaluate the absorption length \abslen,
we compared the PMT outputs in the LXe to those in the GXe
and to a Monte Carlo simulation with no absorption and a variable \raylen.
Note that both absorption and scattering are negligible in the GXe.
In Fig.~\ref{fig:abs} the PMT outputs normalized either
to those in the GXe (a) or
to the MC simulation with \raylen $=45$~cm (b)
are plotted against the source-PMT distance.
Here the distributions at the beginning of purification (open circles)
and after one month of purification (closed circles)
are compared.
The evident exponential decline at the beginning almost vanished
after purification.

These distributions were then fitted with exponential functions.
From the comparison with the GXe data an effective attenuation length
of $54^{+14}_{-9}$~cm was obtained after the purification.
This effective length contains the effects of both
the absorption and the Rayleigh scattering.
Since the scattered photons are not totally lost but may be detected
by other PMTs, the obtained effective attenuation length
is longer than \attlen\ and, especially if \abslen$\gg$\raylen,
it is longer than \raylen.
Note that the LXe/GXe ratio is larger than unity since
the ratio of the scintillation yields in LXe and GXe
is approximately 2.7~\cite{gasw}.

The comparison with the simulation does not show a significant slope.
We used this result to establish a lower limit on the absorption
length \abslen\ of 90~cm at 90~\% confidence level,
where \raylen\ was varied from 30~cm to 50~cm in the simulation.
Before the purification \abslen\ was $12.0\pm 1.8$~cm.
We conclude that the level of water content after the purification
was much lower than 100~ppb.

After successful purification of LXe under the operational
conditions, efforts have been focused on reducing the initial amount
of water contamination in the vessel.
We replaced most of the acrylic parts with Teflon to suppress out-gassing
in the inner chamber, which resulted in an initial absorption length
\abslen\ of 30~cm in the subsequent operation.
A lower limit on the absorption length
\abslen\ of 100~cm at 90~\% confidence level was then achieved
in a much shorter time of xenon purification, in about 300~hours.

We are also developing a liquid-phase purification system with a fluid pump
to increase the purification efficiency.
In such a system the circulation speed could be increased up to
100~$\ell$/hour of LXe, more than 1000 times faster than the current system.

\section{Expected detector performance}
\label{sec:performance}

In this section we estimate the performance of the 100~$\ell$ LXe prototype
for 52.8~MeV $\gamma$ rays that are expected from the \megp decays.

Simulations show that the determination of the incident positions of the
$\gamma$ rays is only slightly affected by absorption and
Rayleigh scattering.
On the contrary the energy resolution is heavily dependent on the
absorption
mainly because of fluctuations in light collection efficiency.
For a short absorption length the total amount of light collected by the PMTs
depends strongly on the positions of the energy deposits.
Although corrections for the first $\gamma$ ray conversion points
may be straightforward,
the total light yield just fluctuates according with the event-by-event
shower fluctuations.
We therefore concentrate on the detector performance in energy measurement
of $\gamma$ rays in the following.

For negligible absorption (i.e. \abslen$\gg$ the detector size)
the $\gamma$ ray energy may be simply evaluated by the total sum
of the photons collected by all the PMTs, possibly weighted by
the local density of the PMTs.
For a finite absorption length, however, a better method of summing the
PMT outputs is necessary.

% (linear fit)
The $\gamma$ ray energy $E$ may be calculated as a linear sum of the
PMT outputs $Q_i$ with arbitrary coefficients $c_i$:
\begin{equation}
E = c + \sum_i c_i Q_i.
\label{eq:linearfit1}
\end{equation}
To optimize the coefficients we may use simulated events with the
$\gamma$ ray energy of $E_t$ and minimize
\begin{equation}
\chi^2 = \left\langle (E-E_t)^2 \right\rangle,
\label{eq:linearfit2}
\end{equation}
where $\langle A \rangle$ is the average of $A$ over the simulated events.
The minimization is straightforward and yields the following result:
\begin{eqnarray}
c_i&=&{\bf M}^{-1}\left(\left\langle E_tQ_i\right\rangle
-\left\langle E_t\right\rangle\bigl\langle Q_i\bigr\rangle\right),
\label{eq:linearfit3}\\
c&=&\langle E_t\rangle -\left\langle\sum_jc_jQ_j\right\rangle.
\label{eq:linearfit4}
\end{eqnarray}
Here ${\bf M}$ is just the covariance matrix of $Q_i$ for the simulated events,
${\bf M}_{kl}\simeq\frac{N}{N-1}\left\langle(Q_k-\langle Q_k\rangle)
(Q_l-\langle Q_l\rangle)\right\rangle$.
This method is called ``linear fit'' and its validity is based on
the principal component analysis~\cite{pcana}.

\begin{figure}[!hbt]
\begin{center}
\epsfig{file=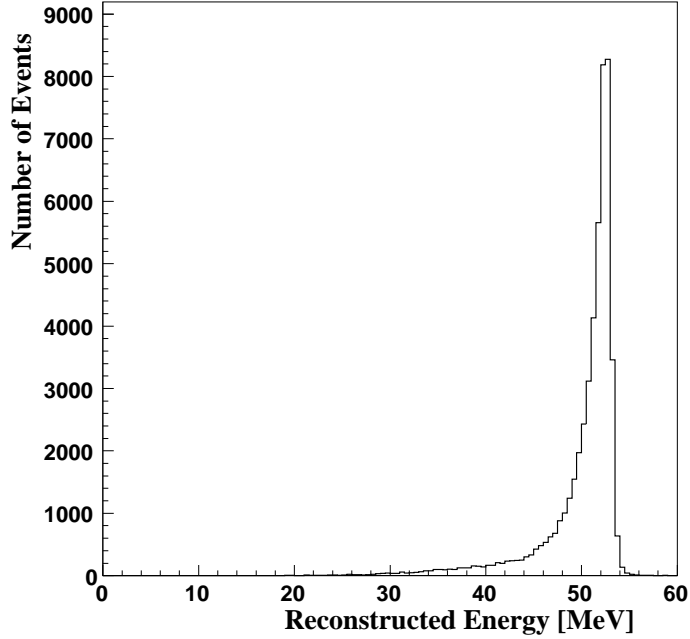,width=9cm}
\end{center}
\caption{A reconstructed energy spectrum for 52.8-MeV $\gamma$ rays
by the linear fit.}
\label{fig:a100hst}
\end{figure}

Using the linear fit,
a reconstructed energy spectrum for 52.8~MeV monochromatic $\gamma$ rays
that uniformly irradiate the center of the detector face
is shown in Fig.~\ref{fig:a100hst}.
The spectrum has an asymmetric shape.  The lower tail is caused by
interactions of the $\gamma$ rays in the materials before the LXe
and by a leakage of shower components (mostly low energy $\gamma$ rays).
Most important for the \megp experiment is the resolution at the
upper edge ($\sigma_\mathrm{u}$) to reject background events,
while the lower tail concerns the detection efficiency.
The spectrum was fitted to the following function
to evaluate $\sigma_\mathrm{u}$:
$$
\begin{array}{cr}
f(E)=\left\{
\begin{array}{ll}
\displaystyle{
\exp\left(\frac{t}{\sigma_\mathrm{u}^2}\left\{\frac{t}{2}-(E-\mu)\right\}\right)},
\qquad&E\leq\mu+t,\\[.3cm]
\displaystyle{
\exp\left\{\frac{(E-\mu)^2}{-2\sigma_\mathrm{u}^2}\right\}},\qquad&E>\mu+t,
\end{array} \right.
\end{array}
$$
where the parameters $\mu$ and $t$ were also determined in the fit.

\begin{figure}[!hbt]
\begin{center}
\epsfig{file=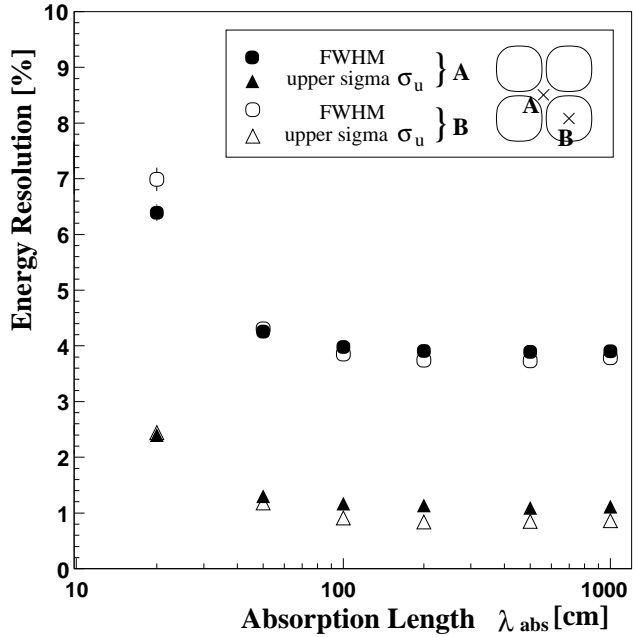,width=0.6\textwidth}
\end{center}
\caption{The expected energy resolutions for 52.8~MeV $\gamma$ rays
as a function of the absorption length.
Resolutions at two incident positions, in the middle of four PMTs (A)
and at the center of one PMT (B), as sketched in the inset,
are compared.}
\label{fig:eresoabs}
\end{figure}

The obtained resolutions in FWHM and $\sigma_\mathrm{u}$
for 52.8~MeV $\gamma$ rays are plotted in
Fig.~\ref{fig:eresoabs} as a function of the absorption lengths
assumed in the simulation.
The resolutions for two incident positions A and B
described in the inset of Fig.~\ref{fig:eresoabs}
are similar, indicating a small position dependence.
The resolutions are also stable by changing \raylen$= 30 - 50$~cm and
for $n = 1.57 - 1.72$ for LXe.
Note that in the simulation refraction, reflection and absorption
of the scintillation light at the PMT quartz windows
are taken into account.

With the achieved absorption length of \abslen$>100$~cm
an energy resolution of 4~\% FWHM
and $\sigma_\mathrm{u}/E \sim 1.2$~\%,
averaged over the detector acceptance, is expected
from the Monte Carlo simulation.
The detection efficiency is estimated to be
approximately 60~\% for \abslen$>50$~cm
if selected within $\pm 4$~\% around the energy peak.

\section{Summary}

We are developing a LXe $\gamma$ ray detector for the MEG experiment.
A 100~$\ell$ prototype with an active volume of
$372\times 372\times 496$~mm$^3$ was constructed to examine its performance
for 52.8-MeV $\gamma$ rays that are expected from \megp decays.
We have established a long stable operation of this new type of device
by successfully running it for approximately 2000~hours
without interruptions.

Absorption of the vacuum ultra violet scintillation light of xenon
by possible impurities might critically limit the performance of the detector.
We developed a method to evaluate absorption of scintillation light in LXe,
separately from Rayleigh scattering, by measuring
cosmic rays and $\alpha$ sources attached inside the detector.
It was found that a ppm level contamination of water is the prime
cause of light absorption in LXe.
By introducing a suitable purification system,
an absorption length longer than 100~cm at 90~\% C.L. has been achieved.

A Monte Carlo simulation study shows that,
with an absorption length of 100~cm or longer,
an energy resolution of 4~\% in FWHM and an upper edge resolution of
$\sigma_\mathrm{u}/E = 1.2$~\% are expected.

To verify the detector performance,
the prototype has been recently irradiated by high energy $\gamma$ rays
from laser Compton back-scatterings
and from $\pi^0$ decays in the charge exchange reactions,
$\pi^- + p \rightarrow \pi^0 + n$.
Analyses of these data are in progress and the results will be
reported elsewhere.

\ack
We wish to thank Counter Hall group of KEK for providing us
with great conveniences
on performing the detector tests.
We also thank Cryogenics group of IPNS, KEK, for supporting us in
operating the cryostat.
The work is supported in part by Grant-in-Aid for Scientific Research
on Priority Areas (A) provided by the Ministry of Education, Culture, Sports,
Science and Technology of Japan.

\end{document}